\documentstyle[12pt,epsfig]{article}
\begin{document}
\begin{center}
{\large{\bf Glueballs and Mesons: the Ground States}}

\vspace*{5mm}

{G.~Ganbold}\footnote{{\tt ~~ganbold@theor.jinr.ru}}

\vspace*{3mm}

{ Bogoliubov Lab. Theor. Phys., JINR, 141980, Dubna, Russia; \\
Institute of Physics and Technology, 210651, Ulaanbaatar, Mongolia}
\end{center}

\begin{abstract}
We provide a new, independent, and analytic estimate of the lowest glueball
mass, and we found it at 1661 MeV within a relativistic quantum-field model
based on analytic confinement.  The conventional mesons and the weak decay
constants are described to extend the consideration. For the spectra of
two-gluon and two-quark bound states we solve the  ladder Bethe-Salpeter
equation. By using a minimal set of model parameters (the quark masses,
the coupling constant, and the confinement scale) we obtain numerical results
which are in reasonable agreement with experimental evidence in the wide
range of energy scale. The model serves a reasonable framework  to describe
simultaneously different sectors in low-energy particle physics.
\end{abstract}

\vskip 5mm


\section{Introduction}

Confinement and dynamical symmetry breaking are two crucial features of QCD,
although they correspond to different energy scales \cite{mira93,alko03}.
Confinement is an explanation of the physics phenomenon that color  charged
particles are not observed; the quarks are confined with other quarks by the
strong interaction to form bound states so that the net color is neutral.
However, there is no analytic proof that QCD should be color confining and
the  reasons for quark confinement may be somewhat complicated. There
exist  different suggestions about the origin of confinement, some dating
back to the  early eighties (e.g., \cite{leut81,stin84}) and some more recent
based on the Wilson loop techniques \cite{kraan98}, string theory quantized
in higher  dimensions \cite{alko07}, and lattice Monte Carlo simulations
(e.g., \cite{lenz04}), etc.  It may be supposed that the confinement is not
obligatory connected with the strong-coupling regime, but it may be induced
by the nontrivial background fields.  One of the earliest suggestion in this
direction is the analytic confinement (AC) based on the assumption that the
QCD vacuum is realized by the self-dual vacuum gluon fields which are stable
versus local quantum fluctuations and related to the confinement and chiral
symmetry breaking \cite{leut81}. This vacuum gluon field could serve as the
true minimum of the QCD effective potential \cite{elis85}. Particularly, it has
been shown that the vacuum of the quark-gluon system has the minimum at
the nonzero  self-dual homogenous background field with constant strength
and the quark and gluon propagators in the background gluon field represent
entire analytic functions on the complex momentum plan $p^2$ \cite{efned95}.
However, direct use of these propagators for low-energy particle physics
problems encounters complex formulae and cumbersome calculations.

We are far from understanding how QCD works at longer distances. The
well-established conventional perturbation theory cannot be used at low energy,
where the most interesting and novel behavior is expected  \cite{pros03}.
The calculations of hadron mass  characteristics on the level of experimental
data precision still remain among the unsolved problems in QCD due to some
technical and conceptual difficulties related  with the color confinement and
spontaneous chiral symmetry breaking.  In such a case, it is useful to
investigate the corresponding low-energy effective theories instead of
tackling the fundamental theory itself.  Although lattice gauge theories are
the way to describe effects in the strong-coupling regime, other methods
can be applied for some problems not yet feasible with lattice techniques.
So data interpretations and calculations of hadron characteristics are
frequently carried out with the help of phenomenological models.  Different
nonperturbative approaches have been proposed to deal with the long
distance properties of QCD, such as  chiral perturbation theory \cite{gass84},
QCD sum rule \cite{shif79}, heavy quark effective theory \cite{neub94}, etc.
Along outstanding advantages these approaches have obvious shortcomings.
Particularly, rigorous lattice QCD simulations  \cite{gupt98} suffer from
lattice artifacts and uncertainties and cannot yet give a reliable result in
the low-energy hadronization region. The coupled Schwinger-Dyson equation
(SDE) is a continuum method without IR and UV cutoffs and describes
successfully the QCD vacuum and the long distance properties of strong
interactions such as confinement and chiral symmetry breaking  (e.g.,
\cite{robe03}). However, an infinite series of equations requires to make
truncations which are  gauge dependent. The  Bethe-Salpeter equation (BSE)
is an important tool for studying the relativistic two-particle bound state
problem in a field theory framework \cite{bethe51}. The BS amplitude in
Minkowski space is singular and therefore, it is usually solved in Euclidean
space to find the binding energy. The solution of the BSE allows to obtain
useful information about the understructure of the hadrons and thus serves
a powerful test for the quark theory of the mesons. Numerical calculations
indicate that the ladder BSE with phenomenological potential models can
give satisfactory results (for a review, see \cite{robe94}).

It represents a certain interest to combine the AC conception and the BSE
method within a phenomenological model and to investigate some low-energy
physics problems by using the path-integral approach. Particularly, it is shown
that a ``toy'' model of interacting  scalar ``quarks''  and ``gluons'' with AC
could result in qualitatively reasonable description of the two- and
three-particle bound states \cite{efim02} and obtained analytic solutions to the
ladder BSE lead to the Regge behaviors of meson spectra \cite{ganb05}.

Below we consider a more realistic model introduced in \cite{ganb05b} by
taking into account the spin, color and flavor degrees of constituents. This
model was further modified in \cite{ganb06a}, applied to leptonic decay
constants in \cite{ganb06b}, and used to simultaneously compute meson masses
and estimate the mass of the lowest-lying glueball in \cite{ganb07,ganb08}.
Here the aim is to collect all necessary formulae, explain the method in detail,
and show that the correct symmetry structure of the quark-gluon interaction in
the confinement region reflected in simple forms of the quark and gluon
propagators can result in quantitatively reasonable estimates of physical
characteristics in low-energy particle physics. In doing so, we build a model
describing hadrons as relativistic bound states of quarks and gluons and
calculate with reasonable accuracy the hadron important characteristics such
as the lowest glueball mass, mass spectra of conventional mesons, and the
decay constants of light mesons.

The paper is organized as follows. In Sec. II we describe the main structure
and specific features of the model. Analytic formulae for the meson spectra
and weak decay constants are derived and numerical results on the vector and
pseudoscalar meson masses and constants $f_\pi$ and $f_K$ are evaluated
in Sec. III.  The formation of a two-gluon bound state, the analytic
expression for the lowest glueball mass and its numerical value are
represented in  Sec. IV.

\section{The Model}

Because of the complexity of QCD, it is often prudent to examine simpler
systems exhibiting similar characteristics first.  Consider a simple relativistic
quantum-field model of quark-gluon interaction assuming that the AC takes
place. The model Lagrangian reads \cite{ganb07}:
\begin{equation}
{\cal L}=-{1\over 4}\left(  F^A_{\mu\nu} -g f^{ABC}{\cal A}^B_\mu
{\cal A}^C_\nu \right)^2 +\sum_{f}\left( \bar{q}^a_f\left[ \gamma_\alpha
\partial^\alpha-m_f+g\Gamma^\alpha_C {\cal A}^C_\alpha
\right]^{ab}q^b_f \right)\,,
\end{equation}
where
${\cal A}^C_\alpha$ -- gluon adjoint representation ($\alpha=\{1,...,4\}$);
$F^A_{\mu\nu}=\partial^\mu {\cal A}_\nu^A-\partial^\nu {\cal A}_\mu^A$;
$f^{ABC}$ -- the $SU_c(3)$ group structure constant ($\{A,B,C\}=\{1,...,8\})$;
$q_f^a$ -- quark spinor of  flavor $f$ with color $a=\{1,2,3\}$ and mass $m_f$;
$g$ -- the coupling strength,
$\Gamma^\alpha_C=i\gamma_\alpha t^C$;
and $t^C$ -- the Gell-Mann matrices.

Consider the partition function
\begin{eqnarray}
Z(g)=\int\!\!\!\int\!\! {\mathcal{D}} \bar{q} ~{\mathcal{D}} q \int\!\!
{\mathcal{D}}  {\cal A} \exp\left\{-\int\!\! dx \,{\cal L}
[\bar{q}, q, {\cal A}] \right\}\,, \qquad Z(0)=1 \,.
\end{eqnarray}
We allow that the coupling remains of order 1 (i.e., $\alpha_s=g^2/4\pi \sim 1$)
in the hadronization region. Then, the consideration may be restricted within
the ladder approximation sufficient to estimate the spectra of two-quark and
two-gluon bound states with reasonable accuracy \cite{ganb06a,ganb07}.
The path integrals defining the leading-order contributions to the two-quark
and two-gluon bound states read:
\begin{eqnarray}
&& Z_{q\bar{q}} = \int\!\!\!\int\!\!{\mathcal{D}}\bar{q}{\mathcal{D}} q
\exp\left\{ -(\bar{q} S^{-1}q)+{g^2\over 2} \left\langle
(\bar{q}\Gamma{\cal A} q )
(\bar{q}\Gamma{\cal A} q ) \right\rangle_D   \right\} \,, \\
&& Z_{{\cal AA}} = \left\langle \exp\left\{-{g\over 2}
\left( f {\cal A} {\cal A} F \right)\right\} \right\rangle_D\,, \qquad
\langle  (\bullet) \rangle_D \doteq \int\!\!{\mathcal{D}} {\cal A}
~e^{-{1\over 2}({\cal A} D^{-1}{\cal A})} (\bullet)\,.
\label{pathint}
\end{eqnarray}

The Green's functions in QCD are tightly connected to confinement and are
ingredients for hadron phenomenology.  The structure of the QCD vacuum
is not well established and one may  encounter difficulties by defining the
explicit quark and gluon propagator  at the confinement scale.  Obviously,
the conventional Dirac and Klein-Gordon forms of the propagators cannot
adequately describe confined quarks and gluons in the hadronization region.
Any widely accepted and rigorous analytic solutions to these propagators are
still missing. Besides, the currents and vertices used to describe the
connection of quarks (and gluons) within hadrons cannot be purely local.
And, the matrix elements of hadron processes are integrated  characteristics
of the propagators and vertices. Therefore, taking into account the correct
global  symmetry properties and their breaking, also by introducing additional
physical parameters, may be more important than the working out in detail
(e.g., \cite{feld00}).

Because of the complexity of explicit Green functions derived in \cite{efned95},
we examine simpler propagators exhibiting similar characteristics.
Consider the  following quark and gluon (in Feynman gauge) propagators:
\begin{eqnarray}
\label{propagat}
&& \tilde{S}_{\pm}^{ab}(\hat{p})
=~\delta^{ab} {i\hat{p}+m_f[1\pm\gamma_5~\omega(m_f/\Lambda)]
\over \Lambda m_f}
\exp\left\{-{p^2+m_f^2\over 2\Lambda^2} \right\}\,,  \nonumber\\
&& \tilde{D}_{\mu\nu}^{AB}(p)=\delta^{AB} {\delta_{\mu\nu}\over p^2}
\exp\left(-{p^2/4\Lambda^2}\right) \,,
\end{eqnarray}
where $\hat{p}=p_\mu \gamma_\mu$ and $\omega(z)=(1+z^2/4)^{-1}$. The
sign ``$\pm$'' in the quark  propagator corresponds to the self- and
antiself-dual modes of the background gluon fields. These propagators are
entire analytic functions in Euclidean space and may serve simple and
reasonable approximations to the explicit propagators  obtained in
\cite{efned95}.  Note, the interaction of the quark spin with  the background
gluon field generates a singular behavior $\tilde{S}_{\pm}(\hat{p})\sim 1/m_f$
in the massless limit $m_f\to 0$. This corresponds to the zero-mode solution
(the lowest Landau level) of the massless Dirac equation in the presence of
external gluon background field and generates a nontrivial quark condensate
$$
 \left\langle \bar{q}_f(0)q_f(0)\right\rangle
= - \int\! {d^4 p\over(2\pi)^4} {\rm Tr} \! \left[\tilde{S}_{\pm}(\hat{p})
\right] =-{6\Lambda^3 \over \pi^2 } \exp\left\{-{m_f^2\over 2\Lambda^2}
\right\} \neq 0
$$
indicating the broken chiral symmetry as $m_f\to 0$.  A mass splitting appears
between  vector and  pseudoscalar mesons ($M_V > M_P$) consisting of  the
same quark content.

Our model has a minimal number of parameters, namely, the coupling constant
$\alpha_s$, the scale of confinement $\Lambda$ and the quark masses
$\{m_{ud},m_s,m_c,m_b\}$. Hereby,  we do not make a distinction of the masses of  lightest
quarks, so $m_u=m_d=m_{ud}$.

Below we describe the main steps in our approach on the example of the
quark-antiquark bound state \cite{ganb08}.

We allocate the one-gluon exchange between colored biquark currents
\begin{eqnarray}
L_2={g^2\over2}\sum\limits_{f_1f_2} \int\!\!\!\int dx_1dx_2
\left( \bar{q}_{f_1}(x_1)i\gamma_\mu t^A q_{f_1}(x_1)\right)
D_{\mu\nu}^{AB}(x_1,x_2)
\left(\bar{q}_{f_2}(x_2)i\gamma_\nu t^B q_{f_2}(x_2)\right).
\label{L2}
\end{eqnarray}

The color-singlet combination is isolated:
$$
(t^A)^{ij}\delta^{AB}(t^B)^{j'i'}={4\over 9} \delta^{ii'}\delta^{jj'}
-{1\over 3}(t^A)^{ii'}(t^A)^{jj'}\,.
$$

We perform a Fierz transformation
\begin{eqnarray*}
(i\gamma_{\mu})\delta^{\mu\nu}(i\gamma_{\nu})
=\sum_{J} C_J \cdot O_J~O_J\,,\qquad J=\{S,P,V,A,T\}\,,
\end{eqnarray*}
where $C_J=\{1,1,1/2,-1/2,0\}$ and $O_J=\{I,i\gamma_{5},
i\gamma_{\mu},\gamma_{5}\gamma_{\mu},  i[\gamma_\mu,\gamma_\nu]/2 \}$.

For systems consisting of quarks with different masses it is important to pass
to the relative co-ordinates $(x,y)$ in the center-of-masses system:
\begin{eqnarray*}
 x_1=x+\xi_1y,~~~~~x_2=x-\xi_2y, \qquad
 \xi_i={m_{f_i}\over m_{f_1}+m_{f_2}}\,,   \qquad i=1,2\,.
\end{eqnarray*}
Then, we rewrite (\ref{L2})
\begin{eqnarray}
L_2={2g^2\over 9}\sum\limits_{Jf_1f_2} C_J \int\!\!\!\int dxdy
 {\cal J}_{Jf_1f_2}(x,y) ~D(y)~{\cal J}^{\dag}_{Jf_1f_2}(x,y),
 \label{Lagran2}
\end{eqnarray}
where
$$
{\cal J}_{Jf_1f_2}(x,y)=\left(\bar{q}_{f_1}(x+\xi_1y)~O_J~q_{f_2}
(x-\xi_2y)\right) \,.
$$

Introduce a system of orthonormalized functions $\{U_Q(x)\}$:
$$
\int\!\! dx \, U_{Q}(x)~U_{Q'}(x)=\delta^{QQ'}\,, \qquad
\sum_{Q} U_{Q}(z) \, U_{Q}(y)=\delta(z-y) \,.
$$

Expand the biquark nonlocal current on the basis
\begin{eqnarray*}
D(y)~{\cal J}^{\dag}_{Jf_1f_2}(x,y) &&=\sqrt{D(y)}
\int dz~ \delta(z-y) \sqrt{D(z)}~{\cal J}^{\dag}_{Jf_1f_2}(x,z) \\
&&= \sum_{Q}\int dz~ \sqrt{D(y)} U_{Q}(y)\cdot
\sqrt{D(z)} U_{Q}(z)~{\cal J}^{\dag}_{Jf_1f_2}(x,z) \,.
\end{eqnarray*}

We define a vertice function $V_{QJ}(x,y)$
$$
\bar{q}_{f_1}(x)~V_{QJ}(x,y)~q_{f_2}(x)
\doteq {2\over 3} \sqrt{C_J} ~\sqrt{D(y)}  U_{Q}(y)
\bar{q}_{f_1}(x+\xi_1y)~O_J~q_{f_2}(x-\xi_2y)
$$
and a colorless biquark current localized at the center of masses:
$$
{\cal J}_{\cal N}(x) \doteq  \int dy ~\left(\bar{q}_{f_1}(x)~V_{QJ}(x,y)
~q_{f_2}(x)\right)  \,, \quad {\cal J}_{\cal N}^{\dag}(x)
={\cal J}_{\cal N}(x)  \,, \quad  {\cal N}=\{QJf_1f_2\} \,.
$$

Then,  (\ref{Lagran2}) can be rewritten as follows:
$$
L_2={ g^2\over 2}\sum\limits_{{\cal N}}\int\!\! dx
\, {\cal J}_{{\cal N}}(x){\cal J}_{{\cal N}}(x) \,.
$$

We represent the exponential by using a Gaussian path integral
$$
e^{{g^2\over2}\sum\limits_{{\cal N}}({\cal J}_{{\cal N}}^2)}
=\left\langle  e^{g(B_{{\cal N}}{\cal J}_{{\cal N}})} \right\rangle_B \,,
\qquad \left\langle (\bullet) \right\rangle_B \doteq
 \int \prod\limits_N {\mathcal{D}}B_N ~e^{-{1\over2}(B_{{\cal N}}^2)}
 (\bullet) \,,  \qquad \left\langle 1 \right\rangle_B=1
$$
by introducing auxiliary meson fields $B_{\cal N}(x)$. Then,
$$
Z_{q\bar{q}} = \left\langle
\int\!\!\!\int\!\!{\mathcal{D}}\bar{q}{\mathcal{D}} q
\exp\left\{ -(\bar{q} S^{-1}q)+ g(B_{{\cal N}}{\cal J}_{{\cal N}})\right\}
\right\rangle_B \,.
$$

Now we can take explicit path integration over quark variables and obtain
$$
Z_{q\bar{q}} \rightarrow Z= \left\langle \exp\left\{ {\rm Tr}
\ln\left[1+g(B_{{\cal N}}V_{{\cal N}})S\right]\right\}  \right\rangle_B \,,
$$
where ${\rm Tr}\doteq {\rm Tr}_c{\rm Tr}_\gamma \sum_{\pm}$; ${\rm Tr}_c$
and ${\rm Tr}_\gamma$ are traces taken on color and spinor indices,
correspondingly, while $\sum_{\pm}$ implies the sum over self-dual and
antiself-dual modes.

\section{Mesons}

In particle accelerators, scientists see ``jets'' of many color-neutral
particles in detectors instead of seeing the individual quarks. This
process is commonly called hadronization and is one of the least
understood processes in particle physics.

We introduce a {\sl hadronization ansatz} and will identify $B_{\cal N}(x)$
fields with mesons carrying quantum  numbers ${\cal N}$. We isolate all
quadratic field configurations ($\sim B^2_{\cal N})$ in the ``kinetic'' term
and rewrite the partition function for mesons \cite{ganb06a}:
\begin{eqnarray}
 Z  = \int {\prod\limits_{\cal N}  {\mathcal{D}}B_{\cal N}  \,\exp
 \left\{ { - \frac{1}{2}  \sum\limits_{{\cal NN'}}  (B_{\cal N} \,
 [\delta^{{\cal NN'}} + \Pi_{{\cal NN'}} ]\,B_{{\cal N'}} )
 - W_{res} [B_{\cal N} ]} \right\}}  \,,
\end{eqnarray}
where the interaction between mesons is described by the residual part
$ W_{res} [B_{\cal N}] \sim 0(B_{\cal N}^3)$.

The leading-order term of the polarization operator is
\begin{eqnarray}
\Pi_{{\cal NN'}}(z_1-z_2)\doteq  \int\!\!\!\int\!\! dx dy \, U_{\cal N}(x)
\, \alpha_s\lambda(z_1-z_2,x,y) \, U_{{\cal N'}}(y)\,,
\end{eqnarray}
where the Fourier transform of the kernel reads
\begin{eqnarray}
\alpha_s\lambda_{{JJ'}}(p,x,y) = \alpha_s\int\!\! dz \, e^{ipz}
\lambda_{JJ'}(z,x,y) =
&& {4 g^2 \sqrt{C_J\,C_{J'}}\over 9} \sqrt{D(x)D(y)}
\int\!\! {d^4 k\over(2\pi)^4}~e^{-ik(x-y)}      \nonumber\\
&& \cdot  {\rm Tr}\left[O_J \tilde{S}\left(\hat{k}+\xi_1\hat{p}
\right) O_{J'} \tilde{S}\left(\hat{k}-\xi_2\hat{p}\right) \right].
\label{Bethe1}
\end{eqnarray}

We diagonalize the polarization kernel on the orthonormal basis $\{U_{\cal N}\}$:
$$
\int\!\! dy \lambda_{{JJ'}}(p,x,y) \, U_{{\cal N}'}(y)
=  \lambda_{{\cal N}}(-p^2)\, U_{{\cal N}'}(x)
$$
or,
$$
\int\!\!\!\int\!\! dx dy U_{{\cal N}}(x) \lambda_{{JJ'}}(p,x,y) U_{{\cal N}'}(y)
=\delta^{{\cal NN}'}~\lambda_{{\cal N}}(-p^2)
$$
that is equivalent to the solution of the corresponding ladder BSE.

In relativistic quantum-field theory  a stable bound state of $n$ massive
particles shows up as a pole in the S-matrix with a center of mass energy.
Accordingly, the meson mass may be derived from the equation:
\begin{eqnarray}
1+\alpha_s\lambda_{{\cal N}}(M_{{\cal N}}^2)=0\,, \qquad -p^2=M_{\cal N}^2 \,.
\label{Bethe2}
\end{eqnarray}

The following renormalization takes place:
\begin{eqnarray}
\label{renormed}
(U_{\cal N}[1+\alpha_s\lambda_{\cal N}(-p^2)]U_{\cal N})
&&=(U_{\cal N}[1+\alpha_s\lambda_{{\cal N}}(M_{{\cal N}}^2)
+\alpha_s\dot\lambda_{\cal N}(M_{\cal N}^2)[p^2+M_{\cal N}^2] U_{\cal N})  \\
&& = (U_R [p^2+M_{\cal N}^2]U_R)\,, \qquad
\dot\lambda_{\cal N}(z) \doteq {d\lambda_{\cal N}(z) \over dz} \,, \nonumber
\end{eqnarray}
where the renormalized state function reads
\begin{eqnarray}
U_R(x)=\sqrt{\alpha_s\dot\lambda_{\cal N}(M_{\cal N}^2)}\cdot U_{\cal N}(x)\,.
\label{renorm}
\end{eqnarray}

The use of the path-integral technique leads to the following practical advantages
over simply solving a BSE with one-boson exchange:

(i) the vacuum functional may be written in alternative representations,
either through original variables of quarks and gluons or, in terms of bound
states, i.e., we obtain so-called ``quark-hadron duality'',

(ii) the  BS kernel (\ref{Bethe1}) is natively obtained in a symmetric form,

(iii) the normalization of the operators of bound states is performed in the most
simple way by keeping the condition $\dot\lambda(M_J)>0$ evident,

(iv) after renormalization (\ref{renormed}) the partition function of the system
of $B_{\cal N}$ fields takes the conventional form with a kinetic term and
interaction parts.

\subsection{Pseudoscalar and vector meson ground states}

In the quark model $\left(q_{f_1}\bar{q}_{f_2}\right)$  bound states are
classified in $J^{PC}$ multiplets. For a pair with spin $s=\{0,1\}$ and
angular momentum $\ell$ the parity is $P=(-1)^{\ell+s}$  and the total spin
is $|\ell-s|<J<|\ell+s|$. Below we consider the meson ground states
($\ell=0,n_r=0$), the pseudoscalar (${\mathbf{P}}: J^{PC}=0^{-+}$) and
vector (${\mathbf{V}}: J^{PC}=1^{--}$) mesons, the most established
sectors of hadron spectroscopy.

We should derive the meson masses from Eq. (\ref{Bethe2}).  The
polarization kernel  $\lambda_{{\cal N}}(-p^2)$ is real and symmetric that
allows us to find a simple  variational solution to this problem. For the
ground state we choose a trial function \cite{ganb06a,ganb08}:
\begin{eqnarray}
\label{testf}
U(x,a)\sim \sqrt{D(x)} \cdot  \exp\left\{-{a\Lambda^2 x^2\over 4}\right\}\,,
\qquad \int\!\! dx \left| U(x,a)\right|^2 =1\,, \qquad a>0\,.
\label{trial1}
\end{eqnarray}

Substituting (\ref{trial1}) into (\ref{Bethe2})  the variational
equation defining the masses of  ${\mathbf{P}}$ and ${\mathbf{V}}$
mesons as follows:
\begin{eqnarray}
1 &=& - \alpha_s\cdot\lambda_J(\Lambda,M_J,m_1,m_2) \nonumber\\
&=& {\alpha_s C_J \Lambda^2 \over 3\pi m_1 m_2}
~\exp\left\{{M^2_J (\xi_1^2+\xi_2^2) -m_1^2-m_2^2 \over 2\Lambda^2} \right\}
\max\limits_{1/4<a<1/2} \left\{ \left[ {(6a-1)(1-2a)\over a}\right]^2
\right. \nonumber\\
&& \!\!\!\!\! \cdot \exp\left[-{a M^2_J(\xi_1-\xi_2)^2\over 2\Lambda^2}\right]
\left[ 4a\rho_J +{M^2_J\over \Lambda^2} \left(\xi_1 \xi_2 + a(2-a\rho_j)
(\xi_1-\xi_2)^2\right)
\right. \nonumber\\
&& \!\!\!\!\! \left. \left. + {m_1 m_2\over\Lambda^2}\left[ 1+\chi_J
~\omega\left({m_1}\right) \omega\left({m_2} \right) \right] \right] \right\}\,,
\label{ground}
\end{eqnarray}
where $C_J=\{1,1/2\}$, $\rho_J=\{1,1/2\}$ and $\chi_J=\{1,-1\}$
for $J=\{P,V\}$.

Localization of the meson field at the center of masses of two quarks
results in the following asymptotic properties. For mesons consisting of
two very heavy quarks ($m_1=m_2=m\gg 1$) we solve (\ref{ground}) and
obtain the correct asymptotic behavior
$$
M_J^2=4m^2+\varepsilon_J\,, \qquad \varepsilon_J \doteq
4 \ln \left( {3\pi \over 32 \, (7-4\sqrt{3}) \, C_J \alpha_s }  \right) \,.
$$
Note, the next-to-leading value $\varepsilon_J $ does not depend on any
masses. Moreover,  $\varepsilon_V > \varepsilon_P$ because the
corresponding Fierz coefficients  obey $C_P=1 > C_V=1/2$. The mass
splitting $M_V > M_P$ remains for ``heavy-heavy'' quarkonia.

For a ``heavy-light'' quarkonium ($m_1\gg 1\,, ~m_2\sim 1$) we estimate the
mass
$$
M_J^2=m_1^2-\epsilon_J\,, \qquad \epsilon_J  \neq  \epsilon_J(M_J) \,.
$$

\subsection{Weak decay constants}

An important quantity in the meson physics is the weak decay constant. The
precise knowledge of its value provides great improvement in our understanding
of various processes convolving meson decays.  For the pseudoscalar mesons
the weak decay constant $f_P$ is defined by the following current-meson duality:
\begin{eqnarray*}
i f_P\,p_\mu
= \langle 0|J_{A}(0)| U_R(p) \rangle\,,
\end{eqnarray*}
where $J_{A}$ is the axial vector part of the weak current and $U_R(p)$ is the
normalized vector of  state.

We estimate
\begin{eqnarray}
 f_P\cdot p_{\mu}  &=& {\sqrt{2}\,g\over 3 }\int\!\!\! {dk\over(2\pi)^4}
 \int\!\! dx \, e^{-ikx} ~U_{R}(x)\sqrt{D(x)}~
{\rm Tr}\left[i\gamma_5 \tilde{S}\left(\hat{k}+\xi_1\hat{p}\right) \,
\gamma_5\gamma_\mu
\tilde{S}\left(\hat{k}-\xi_2\hat{p}\right) \right]   \nonumber\\
&=& p_{\mu} \cdot {32\,\Lambda\, \alpha_s\, \sqrt{2\dot{\lambda}(M^2_P)}
\over 3\, \pi^{3/2} (m_1+m_2)}  {(1-2a_P)(6a_P-1)\over (1+2a_P)^2}
 \left[ 1+ {a_P\over 1+2a_P}{(m_1-m_2)^2\over m_1 m_2}\right]
  \nonumber\\
&& \!\!\!\!\! \cdot
~\exp\left[{M^2_J (\xi_1^2+\xi_2^2) -m_1^2-m_2^2 \over 2}
-{{a_P\over 1+2a_P} M^2_P(\xi_1-\xi_2)^2}\right]  \,,
\label{decay1}
\end{eqnarray}
where $a_P$ is the value of  parameter $a$ calculated for the given meson
with mass $M_{\mathbf{P}}$.

Particularly, for an ``asymmetric'' meson containing an infinitely heavy quark
($m_1\gg m_2 \sim 1$) we obtain the correct asymptotic behavior
$$
f_P\sim 1/\sqrt{m_1}
$$
due to the localization of the meson field at the center of two quark masses.

\subsection{Numerical results}

To calculate the meson masses we need to fix the model parameters. We determine
the quark mass $m_{ud}$ and the coupling constant $\alpha_s$  from equations:
\begin{eqnarray}
1+\alpha_s\lambda_P(\Lambda,138 {\rm \,MeV},m_{ud},m_{ud})=0\,, \qquad
1+\alpha_s\lambda_V(\Lambda,770,m_{ud},m_{ud})=0
\label{fit1}
\end{eqnarray}
by fitting the well-established mesons  $\pi(138)$ and $\rho(770)$ at different
values of $\Lambda$. The remaining constituent quark  masses $m_s, m_c$,
and $m_b$ are determined by fitting the known mesons $K(495)$,
$J/\Psi(3097)$, and  $\Upsilon(9460)$ as follows:
 \begin{eqnarray*}
&& 1+\alpha_s\lambda_P(\Lambda,495,m_{ud},m_s)=0\,, \\
&& 1+\alpha_s\lambda_V(\Lambda,3097,m_c,m_c)=0\,, \\
&& 1+\alpha_s\lambda_V(\Lambda,9460,m_b,m_b)=0\,.
\label{fit2}
\end{eqnarray*}
The dependencies of the estimated  constituent quark masses on $\Lambda$ are
plotted in Fig. 1.

\begin{figure}[thb]
\hskip -15mm
\centerline{\includegraphics[width=80mm,height=80mm]{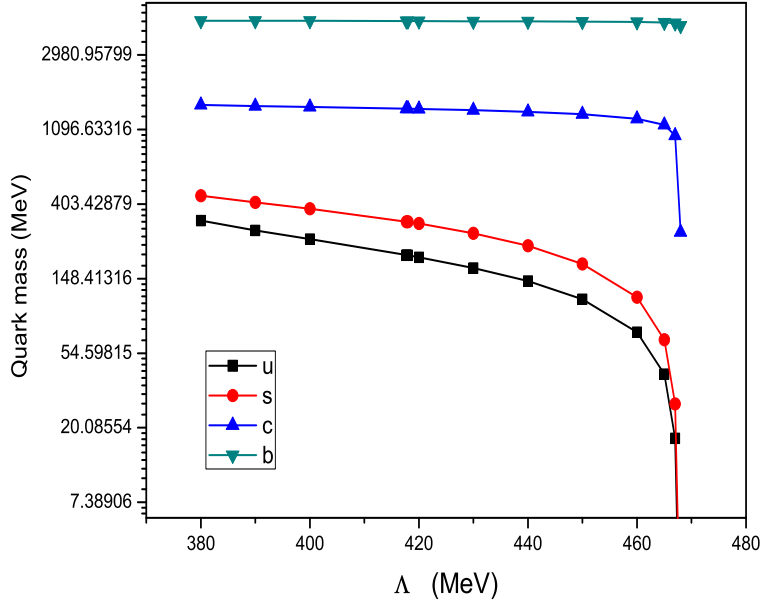}}
\caption{Solutions for constituent quark masses versa the
confinement scale value $\Lambda$.}
\end{figure}

The sharp drop of all quark mass curves in Fig.1 may be shortly explained as
follows. Note, two equations in Eqs. (\ref{fit1}) mostly differ by meson masses
in exponentials along different numerical factors $C_J$, $\rho_J$ and $\chi_J$.
They have general solutions $\{m_{ud}\,,\alpha_s\}$ not for any $\Lambda$.
Suppose, at fixed $\Lambda=\Lambda_0$ they are solvable. Then, for finite
coupling $\alpha_s$  the solution $m_{ud}$ is obviously finite to obey both
equations. However, for vanishing $\alpha_s\to 0$ the equations take the form
$$
1\approx{\alpha_s\, C_J\over m_{ud}^2} \cdot const(\Lambda_0,M_J,\rho_J)
$$
and the solution for quark mass behaves $m_{ud}\sim \sqrt{\alpha_s}\to 0$.
This picture is observed in Fig.1.

By using these quark masses and coupling constant we can estimate other
meson masses in dependence on $\Lambda$ and some results are
shown in Fig. 2.

\begin{figure}[thb]
\hskip -15mm
\centerline{\includegraphics[width=80mm,height=80mm]{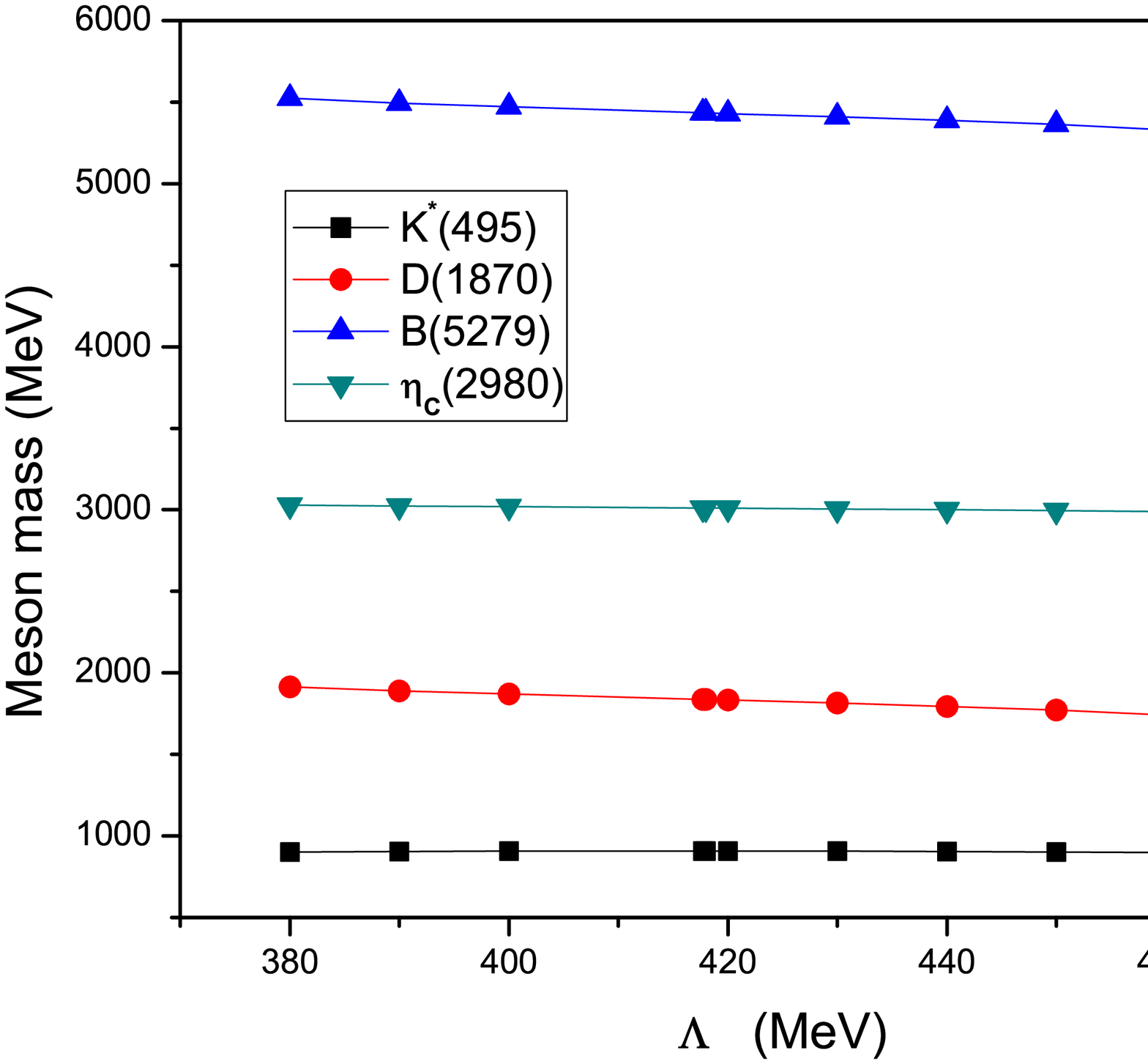}}
\caption{Solutions for some meson masses in dependence on
the confinement scale value $\Lambda$.}
\end{figure}

To fix the value of parameter  $\Lambda$ we calculate the weak decay
constants $f_\pi$ and $f_K$ to compare with experimental data. Note, these
constants considerably depend on $\Lambda$ (see Fig. 3)  that allow us
to fix it unambiguously at $\Lambda=416.4$ MeV.

\begin{figure}[thb]
\hskip -15mm
\centerline{
\includegraphics[width=80mm,height=80mm]{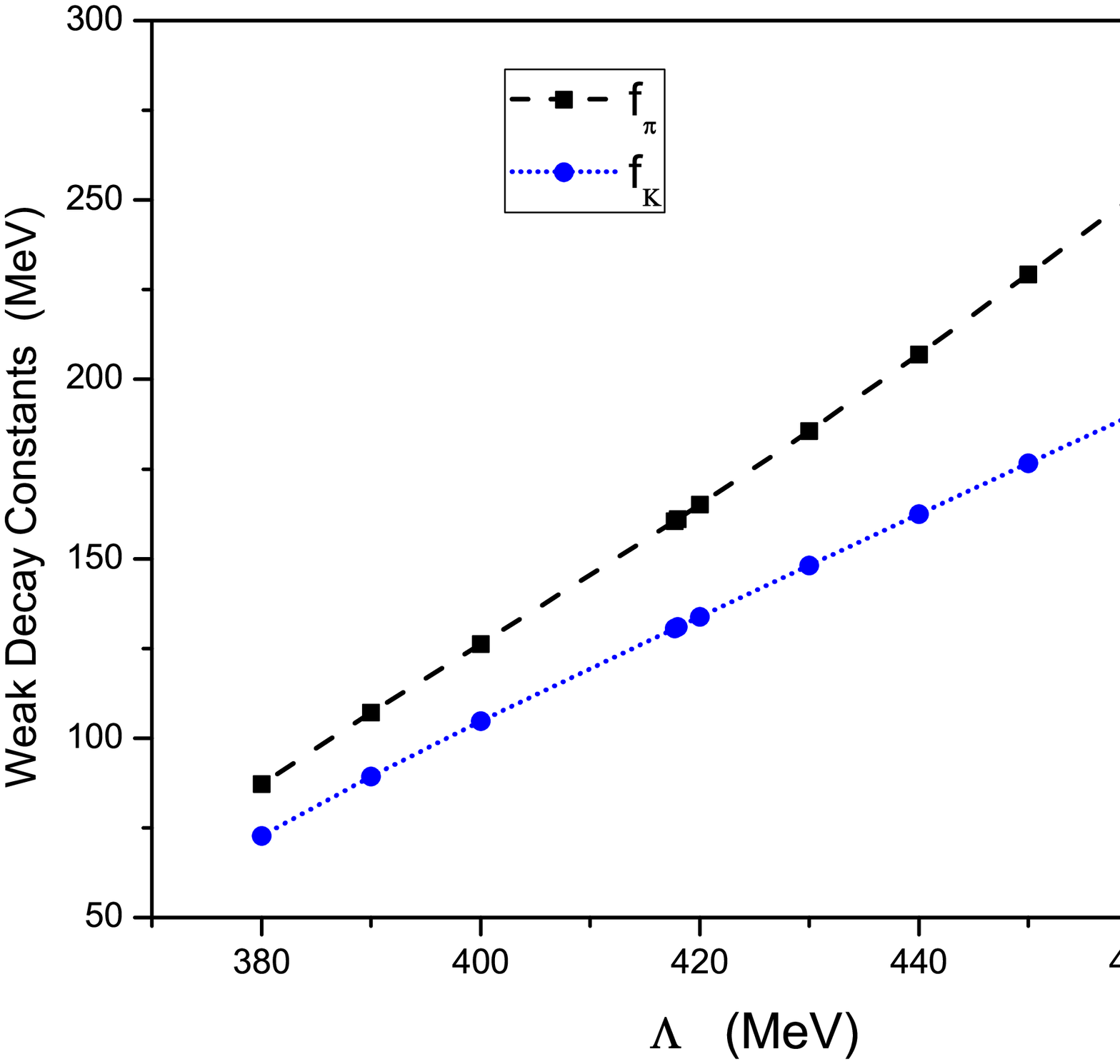} }
\caption{Weak decay constants depending on the confinement
scale value $\Lambda$.}
\end{figure}

The final set of model parameters are fixed as follows:
\begin{eqnarray}
\alpha_s=1.5023 \,,\qquad  \Lambda=416.4 {\mbox{\rm ~MeV}} \,,
\qquad m_{ud}=206.9{\mbox{\rm ~MeV}} \,, \nonumber\\
m_s=323.6{\mbox{\rm ~MeV}}  \,,\qquad m_c=1453.8{\mbox{\rm ~MeV}}  \,,
\qquad m_b=4698.9{\mbox{\rm ~MeV}} \,.
\label{parameters}
\end{eqnarray}
With these parameters we have estimated the pseudoscalar and vector meson
masses shown in Table 1 and compared  with experimental data \cite{PDG2008}.
The relative error of our estimate does not exceed $3.5$ percent in the whole
range of mass (from 0.14 GeV up to 9.5 GeV).

\begin{table}[ht]
\begin{center}
\begin{tabular}{|c|c||c|c||c|c||c|c|}
 \hline
 \hline
$J^{PC}=0^{-+}$ &$M_{\mathbf{P}}$&$J^{PC}=0^{-+}$ & $M_{\mathbf{P}}$
&$J^{PC}=1^{--}$ &$M_{\mathbf{V}}$&$J^{PC}=1^{--}$ & $M_{\mathbf{V}}$\\
 \hline
 $\pi(138)$      & 138   & $\eta_c(2979)$& 3012& $\rho(770)$    & 770
 &$D^*_s(2112)$  & 2078\\
 $K(495)$       & 495   & $B(5279)$     & 5437& $\omega(782)$  & 785
 &$J/\Psi(3097)$ & 3097\\
 $\eta(547)$    & 547  & $B_s(5370)$   & 5551& $K^*(892)$     & 909
 &$B^*(5325)$    & 5464\\
 $D(1870)$     & 1840& $B_c(6286)$   & 6522& $\Phi(1019)$   & 1022
 &$\Upsilon(9460)$&9460\\
 $D_s(1970)$ & 1970& $\eta_b(9300)$& 9434& $D^*(2010)$    & 1942
 &               &     \\
\hline
\hline
\end{tabular}
\caption{Estimated spectrum of conventional mesons
(in units of {\rm MeV}).}
\end{center}
\end{table}

There are mainly two schemes describing $\omega - \Phi$ and $\eta-\eta'$
mixings \cite{PDG2008}.  The octet-singlet scheme uses the mixing angle
$\theta$  between states $(u\bar{u}+d\bar{d}-2 s\bar{s} )/\sqrt{6}$ and
$(u\bar{u}+d\bar{d}+s\bar{s} )/\sqrt{3}$. We use the quark-flavor based
mixing scheme between states $(u\bar{u}+d\bar{d})/\sqrt{2}$ and $s\bar{s}$
with mixing angle $\varphi$. These two schemes are equivalent to each other
by $\theta=\varphi-\pi/2+\arctan(1/\sqrt{2})$ when the $SU(3)$ symmetry is
perfect. Particularly, for ``ideal'' vector mixing the angle is
$\varphi_V^{id}=90^\circ$ or  $\theta_V^{id}=35.3^\circ$.

With fixed parameters (\ref{parameters}) we calculate a relatively heavy mass
$M_V(s\bar{s})=1064$ MeV of vector $s\bar{s}$ state. To obtain correct masses
of $\omega(782)$ and $\Phi(1019)$ one needs a considerable mixing to the light
quark-antiquark state with mixing angle $\varphi_V\simeq 73.2^\circ$ which
differs significantly from the ``ideal'' value. By using the same parameters
(\ref{parameters}) we obtain a pseudoscalar $s\bar{s}$ state with mass
$M_P(s\bar{s})=705$ MeV.  We cannot describe the physical mass of
$\eta'(958)$ by any mixing to the light-quark pair and can only fit the correct
mass $M_P(\eta)=547$ MeV at angle $\varphi_P\simeq 58.5^\circ$.  Our model
fails to describe simultaneously the $\eta - \eta'$ mixing. This problem obviously
deserves a separate consideration.

Note, the infrared behavior of effective (mass-dependent) QCD  coupling
$\alpha_s$ is not well defined and needs to be more specified
\cite{shir02,nest03,kasz05}. In the region below the $\tau$-lepton mass
($M_\tau=1.777$ GeV) the strong-coupling value is expected between
$\alpha_s(M_\tau) \approx 0.34$ \cite{PDG2008} and the infrared fix point
$\alpha_s(0)=2.972$ \cite{fisch05}.  Our parameter $\alpha_s=1.5023$ does
not contradict this expectation because it is estimated to fit the $\pi$ meson
mass, and so the corresponding energy scale is $\sim 140$ MeV.
We keep this value for further calculations.

The weak decay constants of light mesons are well established data and
many groups (MILC \cite{bern07}, NPLQCD \cite{bean07}, HPQCD \cite{foll07},
etc.) have these with accuracy at the
2 percent level. Therefore, these values are often used to test any model
in QCD.  By substituting optimal values of $\{m_{ud}, m_s, \alpha_s, \Lambda\}$
(\ref{parameters}) into (\ref{decay1}) we calculate
$$
f_\pi=128.8 {~\mbox{\rm MeV}} \,, \qquad f_K=157.7 {~\mbox{\rm MeV}} \,.
$$
Our estimates are in agreement with the experimental
data \cite{bern05,PDG2008}:
\begin{eqnarray}
f_{\pi^-}^{PDG}=130.4\pm 0.04 \pm 0.2 \, {\rm MeV}\,, \qquad
f_{K^-}^{PDG}=155.5\pm 0.2 \pm 0.8 \pm 0.2 \, {\rm MeV}\,.
\label{decay3}
\end{eqnarray}

Our model represents a reasonable framework to describe the conventional
mesons, and the parameters are fixed. Below we can consider two-gluon
bound states.

\section{Glueball Lowest State}

Because of the confinement, gluons are not observed, they may only come
in bound states called {\sl glueballs}. Glueballs are the most unusual particles
predicted by the QCD but not found experimentally yet \cite{klem07}.  There
are predictions expecting  non-$q\bar{q}$ scalar objects, like glueballs and
multiquark states in the mass range $\sim 1500 - 1800{\mbox{\rm ~MeV}}$
\cite{amsl04,bugg04}. Experimentally the closest scalar resonances to this
energy range are the $f_0$(1500) and $f_0$(1710) \cite{yao06}.
Some references favor the $f_0$(1500) as the lightest scalar glueball
\cite{bugg00}, while others do so for the $f_0$(1710) \cite{sext95,chan07}.
Recent scalar hadron $f_0$(1810) reported  by the BES collaboration
may also be a glueball candidate \cite{abli06}.

The study of glueballs currently deserves much interest from a theoretical
point of view, either within the framework of  effective models or lattice QCD.
The glueball spectrum has  been studied by using effective approaches
like the QCD sum rules \cite{nari00},  Coulomb gauge QCD  \cite{szcz03},
and potential models (e.g., \cite{corn83,brau04}), etc. The potential models
consider glueballs as bound states of two or more constituent gluons
interacting via a phenomenological potential \cite{corn83,kaid06,math06}.
It should be noted that  potential models have difficulties in reproducing all
known lattice QCD data.  Different string models are used for describing
glueballs \cite{yama77,solo01}, including  combinations of string and
potential approaches \cite{brau04}.  It has been shown that a proper inclusion
of the  helicity degrees of freedom can improve the compatibility between
lattice QCD and potential models  \cite{math06a}.

An important theoretical achievement in this field  has been the prediction
and computation of the glueball spectrum in lattice QCD simulations
\cite{morn99,meye05}. Recent lattice calculations, QCD sum rules, "tube"
and constituent glue models predict that the lightest glueball has the quantum
numbers of scalar ($J^{PC}=0^{++}$) and tensor ($2^{++}$) states
\cite{ochs06}. Gluodynamics  has been extensively investigated within
quenched lattice QCD simulations and the lightest glueball is found
a scalar object with a mass of $\simeq 1.66\pm 0.05$ GeV \cite{vacc99}.
A use of much finer isotropic lattices resulted in a value 1.475 GeV
\cite{meye05}. Recently, an improved quenched lattice calculation of the
glueball spectrum at the infinite volume and continuum limits based on
much larger and finer lattices have been carried out  and the scalar glueball
mass is calculated to be  $1710\pm50\pm80$ MeV \cite{chen06}.

Two-gluon bound states are the most studied purely gluonic systems
in the literature, because when the spin-orbital interaction is ignored
($\ell =0$), only scalar and tensor states are allowed.
Particularly,  the lightest glueballs with positive charge parity can be
successfully modeled by a two-gluon system in which the constituent
gluons are massless helicity-one particles \cite{math08}.

Below we consider a two-gluon scalar bound state.  We isolate the
color-singlet term in the bi-gluon current in $Z_{{\cal AA}}$
(\ref{pathint}) by using the known relations
\begin{eqnarray*}
&&
t^C_{ik}\,t^C_{jl}={N_c^2-1\over 2N_c^2} \delta^{il}\delta^{jk}
-{1\over N_c} t^C_{il}t^C_{jk}\,, \\
&&
f^{ABE}f^{A'B'E}={2\over 3}\left(\delta^{AA'}\delta^{BB'}
-\delta^{AB'}\delta^{BA'}\right) + d^{AA'E}d^{BB'E} -d^{AB'E}d^{BA'E}\,.
\end{eqnarray*}
The second-order matrix element  containing a color-singlet two-gluon
current reads \cite{ganb07}
\begin{eqnarray*}
L_{{\cal AA}}=
&&{g^2 \over  4\cdot 3} \int\!\!\!\int dx dy \left( J^{AA}_{\mu\mu'}(x,y)
J^{BB}_{\nu\nu'}(x,y)-J^{AA}_{\mu\nu'}(x,y) J^{BB}_{\nu\mu'}(x,y)
\right) \\
&& \cdot \left[
\delta^{\nu\nu'} W_{\mu\mu'}(x,y)
-\delta^{\mu\nu'} W_{\nu\mu'}(x,y)
-\delta^{\nu\mu'} W_{\mu\nu'}(x,y)
+\delta^{\mu\mu'} W_{\nu\nu'}(x,y)
\right] \,,
\end{eqnarray*}
where
$$
J^{BC}_{\mu\nu}(x,y) \doteq {\cal A}^B_\mu(x) {\cal A}^C_\nu(y)\,,
$$
$$
W_{\mu\nu}(x,y)\doteq {\partial \over \partial x_\mu}
{\partial \over \partial y_\nu}D(x-y)=\delta^{\mu\nu}\, W(x-y)+\ldots\,,
\qquad  W(z)={1\over (2\pi)^2}\, e^{-z^2} \,.
$$
This part consists of  spin-zero (scalar) and spin-two (tensor) components.
Below we consider the scalar component:
\begin{eqnarray*}
L_{{\cal AA}}^S=&&{g^2 \over 3} \int\!\!\!\int dx_1 dx_2
J(x_1,x_2) W(x_1-x_2)J(x_1,x_2) \,, \quad  J(x_1,x_2)
\doteq J^{BB}_{\mu\mu}(x_1,x_2) \,.
\end{eqnarray*}

By introducing  the relative coordinates
($x_1 \doteq x+y/2\,,~~ x_2 \doteq x-y/2$) we rewrite
\begin{eqnarray}
L_{{\cal AA}}^S=&&{g^2 \over 3} \int\!\!\!\int\!\! dx dy\,
J(x,y) W(y) J(x,y) \,.
 \label{Lagran2a}
\end{eqnarray}

One can see that the matrix element (\ref{Lagran2a}) is similar to
(\ref{Lagran2}) by the very construction. By omitting details of
intermediate calculations (similar to those represented in the previous
section) we  rewrite the partition function in terms of  auxiliary field
$B(x)$ as follows:
$$
Z_{{\cal AA}} \rightarrow Z_G= \int {\mathcal{D}}B
\exp\left\{ -{1\over2}\left(B\, G^{-1} B\right)+L_I[B] \right\} \,,
$$
where $L_I[B]\sim O(B^3)$ and the BS kernel is
\begin{eqnarray*}
&& G^{-1}(x-y)=\delta(x-y)-{8\, g^2\over 3} \Pi(x-y) \,, \\
&& \Pi(z)\doteq \int\!\!\!\int\!\! dt ds ~U_n(t)  \sqrt{W(t)}
~D\left({t+s \over 2}+z\right)D\left({t+s \over 2}-z\right)
\sqrt{W(s)} ~U_n(s)\,.
\end{eqnarray*}

The hadronization ansatz allows us to identify $B$ with scalar glueball
field. To find the glueball mass we should diagonalize the Bethe-Salpeter
kernel $\Pi(z)$. The glueball mass $M_G$ is defined from equation
\cite{ganb08}:
\begin{eqnarray}
1- {8\,g^2\over 3} \int\! dz \, e^{izp} ~\Pi(z)=0 \,, \qquad  p^2=-M_G^2 \,.
\label{gluemass}
\end{eqnarray}

For the lightest ground-state scalar glueball choose a Gaussian wave function:
$$
U(x)={2c \over \pi}~ e^{-c x^2} \,,  \qquad \int\!\!dx \, |U(x)|^2=1\,,
\qquad c>0\,.
$$
Then, we derive (\ref{gluemass}) as follows:
$$
1={ \alpha_s \over \alpha_{crit}} \exp\left\{ {M_G^2\over 4\Lambda^2}
\right\}\,, \qquad \alpha_{crit} \doteq {3\pi(3+2\sqrt{2})^2\over 4} \,.
$$

The final analytic result for the lowest-state glueball mass reads
\begin{eqnarray}
\label{glueball}
M_G = 2\Lambda \left[ \ln\left( {\alpha_{crit}\over \alpha_s} \right)
\right]^{1/2}\,.
\label{gluemass2}
\end{eqnarray}
The solution $M_G^2\ge 0$ exists for any
$\alpha_s < \alpha_{crit}\approx 80.041$.

Note, the scalar glueball mass depends linearly on the confinement scale
$\Lambda$ and the scaled mass $M_G/\Lambda$ depends only on coupling
$\alpha_s$ (see Fig. 4).   Particularly, if  we take values
$\Lambda \sim \Lambda_{QCD}\approx 360$ MeV and
$\alpha_s \simeq \alpha_s(M_\tau)=0.343$, then we estimate
$M_G\approx 1710$ MeV.

However, our purpose is to describe simultaneously different sectors of
low-energy particle physics.  Accordingly, with values $\alpha_s=1.5023$ and
$\Lambda=416.4{\rm \,MeV}$  determined by fitting the meson masses and
weak decay constants, we calculate the scalar  glueball mass as follows:
\begin{eqnarray}
M_G = 1661{\mbox{\rm ~MeV}} \,.
\label{Glueballmass}
\end{eqnarray}
Our estimate (\ref{Glueballmass}) is in reasonable agreement with other
predictions expecting the lightest glueball located in the scalar channel
in the mass range $\sim 1500 - 1800{\rm ~MeV}$
\cite{amsl04,nari00,meye05,bali01}. The often referred quenched QCD
calculations predict $1750 \pm 50 \pm 80{\mbox{\rm ~MeV}}$  for the
mass of the lightest glueball \cite{morn99}. The recent quenched lattice
estimate with improved lattice spacing favors a  scalar glueball mass
$M_G=1710\pm 50 \pm 58 {\rm ~MeV}$ \cite{chen06}.

\begin{figure}[thb]
\centerline{
\includegraphics[width=80mm,height=80mm]{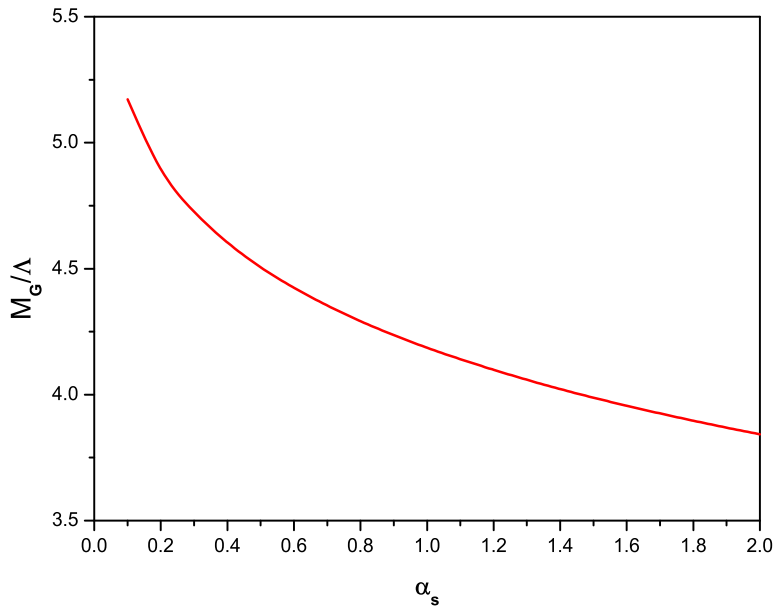} }
\caption{Evolution of the lowest-state glueball mass scaled to
$\Lambda$  with the coupling $\alpha_s$.}
\end{figure}

Another important  property of the scalar glueball is its size, the ``radius''
which should depend somehow on the glueball mass.  We estimate the
glueball size by using the ``effective potential'' $W(y)$ (\ref{Lagran2a})
connecting  two scalar gluon currents.  The glueball radius may be roughly
estimated as follows
\begin{eqnarray}
r_{G} \sim \sqrt{ {\int\! d^4 x~x^2~W(x) \over\int\! d^4 x ~W(x)}}
={\sqrt{2}\over\Lambda} \approx {1\over 295~\mbox{\rm MeV}}
\approx  0.67~\mbox{\rm fm} \,.
\label{radius}
\end{eqnarray}
This means that  the dominant forces responsible for binding gluons must
be provided by medium-sized vacuum fluctuations of correlation length
$\sim 0.7\,\mbox{\rm fm}$. Consequently, typical energy-momentum transfers
inside a scalar glueball occur at the QCD scale $\sim 360$ MeV, rather than
at the chiral symmetry breaking scale $\Lambda_\chi \sim 1$ GeV
(or, $\sim 5\, {\rm fm}$).

From (\ref{gluemass2}) and (\ref{radius}) we deduce that
$$
r_{G} \cdot M_G
=2\sqrt{2}\left[ \ln\left( {\alpha_{crit}\over \alpha_s} \right) \right]^{1/2}
\approx 5.64 \,.
$$
This value may be compared with the prediction
($r_{G} \cdot M_G = 4.16\pm 0.15$) of quenched QCD calculations
\cite{morn99,chen06}.  A study of the glueball  properties at finite
temperature using SU(3) lattice QCD at the quenched level  with the
anisotropic lattice imposes restrictions on the glueball  parameters at
zero temperature:
$0.37\, {\rm fm} < r_G < 0.57\, {\rm fm}$ and  $M_G\simeq1.49\, {\rm GeV}$
 \cite{iish01}.  The nonprincipal differences of quenched lattice QCD data
 from our  estimates may be explained by the presence of quarks
 (our parameters have been  fixed by fitting two-quark bound states) in
 our model.

A method of analysis of correlation functions in QCD is to calculate the
corresponding condensates. The value of the correlation function dictates
the values of the condensates.  We calculate the lowest  nonvanishing
gluon condensate in the leading-order (ladder) approximation:
$$
g^2 Tr \left\langle F_{\mu\nu}^A  F^{\mu\nu}_A  \right\rangle
=8 N_c \pi\alpha_s \Lambda^4  \int\!\! d^4 z \, W(z)=6\pi\alpha_s\Lambda^4
\approx 0.8~ GeV^4
$$
which is the same order of magnitude with the reference value \cite{shif98}
$$
g^2 Tr \left\langle G_{\mu\nu}  G^{\mu\nu} \right\rangle
\approx 0.5~ GeV^4 \,.
$$

\vskip 5mm

In conclusion, the suggested model in its simple form is far from real QCD.
However, our aim is to demonstrate that global properties of the lowest
glueball state and conventional mesons may be explained in a simple way
in the framework of a simple relativistic quantum-field model of quark-gluon
interaction based on analytic confinement. Our guess about the symmetry
structure of the quark-gluon interaction in the confinement region has been
tested and the use of simple forms of propagators has resulted in
quantitatively reasonable estimates in different sectors of the low-energy
particle physics. The consideration can be extended to other problems
in hadron physics.

\vskip 5mm

The author thanks  G.V.~Efimov,  E.~Klempt and  V.~Mathiew for  valuable
remarks and useful suggestions.


\end{document}